\documentclass[aps,prb,showpacs,preprint,amsmath,amssymb,superscriptaddress]{revtex4}

\usepackage{graphicx}
\usepackage{dcolumn}
\usepackage{bm}
\usepackage{color}

\begin{document}

\preprint{APS/123-QED}

\title{Theoretical Analysis of Electronic and Magnetic Properties of NaV$_2$O$_4$: Crucial Role of the Orbital Degrees of Freedom}

\author{Z. V. Pchelkina}
\email{pzv@ifmlrs.uran.ru}
\affiliation{Institute of Metal Physics, Russian Academy of
Sciences-Ural Division, 620990 Ekaterinburg, Russia}
\affiliation{Theoretical Physics and Applied Mathematics
 Department, Ural Federal University, Mira Str. 19, 620002 Ekaterinburg, Russia}

\author{ I. V. Solovyev }
\email{SOLOVYEV.Igor@nims.go.jp}
\affiliation{Computational Materials Science Unit, National Institute for Materials Science, 1-2-1 Sengen, Tsukuba, Ibaraki 305-0047, Japan}

\author{ R. Arita }
\affiliation{Department of Applied Physics, Graduate School of Engineering, The University of Tokyo, 7-3-1 Hongo, Bunkyo-ku, Tokyo 113-8656, Japan}
\affiliation{PREST, Japan Science and Technology Agency, 4-1-8 Honcho, Kawaguchi, Saitama 332-0012, Japan}
\date{\today}

\begin{abstract}
Using realistic low-energy model with parameters derived from the first-principles
electronic structure calculation, we address the origin of the quasi-one-dimensional behavior in
orthorhombic NaV$_2$O$_4$, consisting of
the double chains of edge-sharing VO$_6$ octahedra. We argue that the geometrical aspect alone
does not explain the
experimentally observed anisotropy of electronic and magnetic properties of NaV$_2$O$_4$.
Instead, we attribute the unique behavior of NaV$_2$O$_4$ to one particular type of the orbital ordering,
which respects the orthorhombic $Pnma$ symmetry. This orbital ordering acts to divide all
$t_{2g}$ states into two types: the `localized' ones, which are antisymmetric with respect to the
mirror reflection $y \rightarrow -$$y$, and the symmetric `delocalized' ones. Thus, NaV$_2$O$_4$
can be classified as the double exchange system.
The directional orientation of symmetric orbitals, which form the metallic band,
appears to be sufficient to explain both
quasi-one-dimensional character of interatomic
magnetic interactions and the anisotropy of electrical resistivity.
\end{abstract}

\maketitle

\section{\label{Intro} Introduction}

  Double exchange (DE) is the key concept in the physics of strongly correlated systems,
characterizing the properties of itinerant electrons, traveling in the lattice of
localized (atomic) spins.\cite{DE_oldies}

  The situation becomes increasingly interesting when there are several degenerate orbitals,
which become involved into the DE processes. Typically, each orbital is highly nonspherical and
can assist the electron transfer only in some spacial directions, which can differ substantially,
depending on the shape of the occupied orbital. Thus, the itinerant electron can ``choose'' the
orbital where to reside and, depending on it, choose a path for its traveling in the solid.
Of course, the transport properties will crucially depend on such a choice.
Moreover, each guess for the orbital configuration has a feedback effect and
involves some elements of the self-organization, typically through developing certain
magnetic texture, which tends to stabilize the assumed orbital configuration, at least locally.
Alternatively,
each change of the magnetic state
leads to the orbital-selective reconstruction of the electronic structure, which can be sufficient to
stabilize the magnetic state.\cite{DE_degenerate} All these phenomena are largely involved
in the physics of pseudocubic perovskite manganites and predetermine the rich variety of their electronic
and magnetic properties.\cite{ManganitesReviews}

\begin{figure}[h!]
\begin{center}
\resizebox{8cm}{!}{\includegraphics{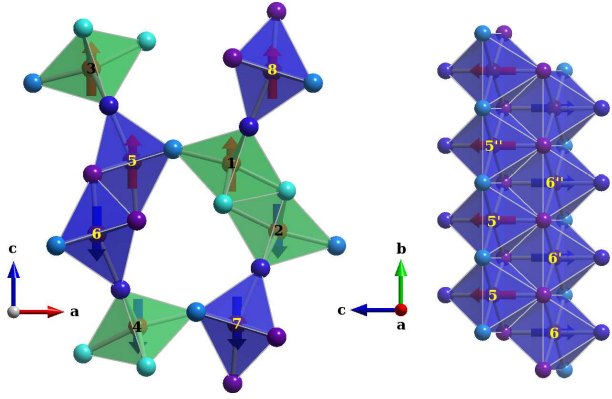}}
\end{center}
\caption{\label{afm} (color online).
The schematic view on the crystal structure and the AFM arrangement of eight vanadium atoms in the primitive cell of NaV$_2$O$_4$.
Eight vanadium atoms in the primitive cell are
in the centers of octahedra and denoted by numbers (1-4 correspond to the type V1 and
5-8 -- to the type V2). Small spheres in the corners of octahedra stand for the nonequivalent oxygen atoms. 
Na atoms are hidden for clarity.
}
\end{figure}

  In this paper we propose that the DE physics can lead to a number of interesting effects when it brought
in contact with the unusual crystal structure of recently synthesized mixed-valence ($d^{1.5}$)
NaV$_2$O$_4$ compound consisting of double chains of edge-sharing VO$_6$ octahedra
(see Fig.~\ref{afm}).\cite{yamaura_07,takagi}

  The properties of NaV$_2$O$_4$ are indeed very intriguing.
It has orthorhombic space group $Pnma$, where
there are four formula units in the primitive cell, and two nonequivalent vanadium types V1 and V2,
which are numbered as 1-4 and 5-8, respectively (Fig.~\ref{afm}).\cite{takagi}
The vanadium atoms are surrounded by slightly distorted edge-sharing octahedra which form the zig-zag
chains propagating in the $b$-direction.

The single-crystalline NaV$_2$O$_4$ remains metallic down to 40mK.
Nevertheless, the low-temperature
properties of NaV$_2$O$_4$ are characterized by the strong anisotropy of the electrical resistivity:
$\rho_\perp$/$\rho_\parallel>$20, where the symbols $\parallel$ and $\perp$ stand for the
directions being parallel and perpendicular to the crystallographic $b$-axis, respectively.
In this sense, NaV$_2$O$_4$ is regarded as the quasi-one-dimensional (quasi-1D) compound.

The magnetic susceptibility ($\chi$) of NaV$_2$O$_4$ has a peak at around $T_N$=140 K,
indicating at an antiferromagnetic (AFM) transition.
On the other hand, the fitting of $\chi$ in terms of the Curie-Weiss law well above $T_N$
yields positive Weiss constant, indicating at the predominantly
ferromagnetic (FM) character of interactions.\cite{yamaura_07}
Furthermore, the anisotropy of electrical resistivity rapidly deteriorates above $T_N$.\cite{yamaura_07,takagi}
These results, together with
the metallic character of conductivity along the $b$-axis gave rise to
a hypothesis that within each zig-zag double-chain, the intrachain interaction
should be FM while the interchain interaction is AFM (Fig.~\ref{afm}).\cite{yamaura_07}
Thus, it seems that not only the electrical resistivity, but also the magnetic
structure of NaV$_2$O$_4$ is highly anisotropic and these two anisotropies are coupled to each other.

  Later on,
several attempts to clarify the magnetic structure of NaV$_2$O$_4$
have been undertaken in the experiments on the neutron scattering,\cite{neutron}
the muon-spin rotation,\cite{muon} and the nuclear magnetic resonance (NMR).\cite{takeda_11}
Nevertheless, the magnetic structure remains a matter of controversy. For example, earlier
neutron-scattering measurements have been interpreted in terms of the incommensurate
spin-density wave order with ${\bf q} = (0,0.191,0)$, in units of $2\pi/b$.\cite{neutron}
However, more resent muon-spin rotation studies favor
the helical magnetic structure, where the magnetic moments lie in the $ac$ plane and propagate
along the $b$ axis. This means that some of the intrachain interactions should be AFM,
which confronts the hypothesis based on the magnetization measurements.\cite{yamaura_07}
Finally, there are also experimental proposals that NaV$_2$O$_4$ may exhibit
several consequent AFM transitions below $T_N$.\cite{muon,takeda_11,sakurai}

  The purpose of this work is to build some theoretical background for understanding
these electronic and magnetic properties of NaV$_2$O$_4$.
First, we construct a realistic low-energy model (Sec.~\ref{Method}), which
aims to describe the electronic and magnetic properties of NaV$_2$O$_4$.
Then, we analyze this model in the mean-field Hartree-Fock (HF) approximation (Sec.~\ref{Results}).
Although HF is a crude approximation for metallic systems, we will be able to argue,
on a semi-quantitative level, that the main details of the electronic and magnetic structure
are well anticipated from such a model analysis and can be attributed to the
particular type of the orbital ordering in the metallic state. Once the orbital ordering
is established, NaV$_2$O$_4$ has many things in common with the DE systems.
Finally, a brief summary will be given in Sec.~\ref{summary}.

\section{\label{Method} Method}

 In our theoretical analysis, we follow the strategy, which was developed
in the previous publications and which can be called
as ``realistic modeling of strongly correlated systems'' (see Ref.~\onlinecite{downfolding} for a review). First, we calculate the
electronic structure of NaV$_2$O$_4$ in the local-density-approximation (LDA),
using the experimental parameters of the crystal structure, measured at
$T$$=$$300$ K.\cite{takagi,remark2} The obtained electronic structure is in a good agreement with
the previous calculations.\cite{yamaura_07}
Then, we construct the low-energy Hubbard-type model for the
$t_{2g}$ bands of NaV$_2$O$_4$, located near the Fermi level,
and derive all parameters of such model from the first-principles electronic structure calculations.
The details of the computational procedure can be found in Ref.~\onlinecite{downfolding}.
In a number of cases, we have also constructed and solved the five-orbital model,
which describes the behavior of both $t_{2g}$ and $e_g$ bands of NaV$_2$O$_4$.

  The model includes three sets of parameters: the crystal field (CF), transfer integrals, and screened Coulomb interactions.

  The
splitting of three $t_{2g}$ levels by the CF has the following structure (in meV):
$(0,46,204)$ and $(0,10,264)$, at the inequivalent vanadium types V1 and V2, respectively.
Two lowest levels are nearly degenerate and not disrupted by the crystal distortion.
This type of the CF splitting is favored by the DE interactions and minimizes the
kinetic energy of electrons.
Nevertheless, two lower $t_{2g}$ orbitals can still mix
with the upper one by the transfer integrals. This mixing can be relatively strong.

\begin{figure}[h!]
\begin{center}
\resizebox{8cm}{!}{\includegraphics{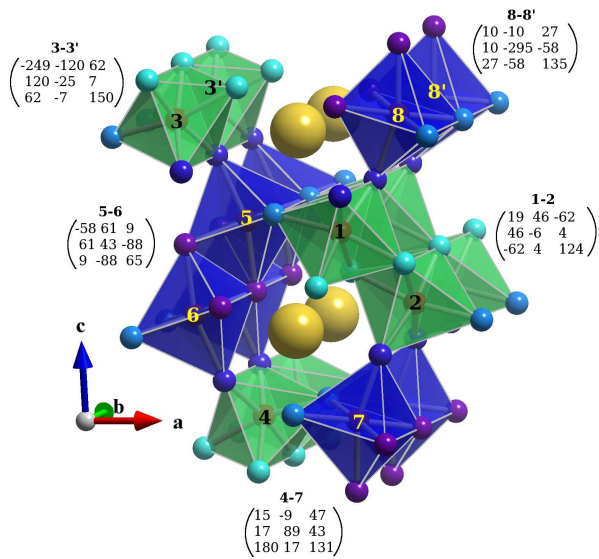}}
\end{center}
\caption{\label{struct} (color online).
Fragment of the crystal structure of NaV$_2$O$_4$ with the matrices of transfer integrals
between nearest neighbors. Each matrix is computed in the
crystal field coordination frame, in the basis of $t_{2g}$ states.
The matrix elements are measured in meV. The big yellow spheres stand for Na atoms. 
Other notations are the same as in Fig.~\ref{afm}.}
\end{figure}

  The behavior of some transfer integrals is explained in Fig.~\ref{struct},
in the local coordinate frame, which
diagonalizes the CF Hamiltonian.
The largest transfer integrals operate between nearest-neighbor (NN) V-atoms
in the chains $1$-$1'$ ($3$-$3'$) and $5$-$5'$ ($8$-$8'$). The transfer integrals between the chains are somewhat weaker,
but still comparable with the ones within the chains. Thus, the structure of
transfer integrals in is essentially three-dimensional, and alone does not explains the quasi-1D character of
the electronic and magnetic properties of NaV$_2$O$_4$.

  The screened Coulomb interactions in the $t_{2g}$-band can be computed in two steps,\cite{downfolding} by applying
the constrained LDA and the random-phase approximation (RPA) for the screening. Roughly speaking,
the first techniques takes into account the screening of atomic orbitals, while the second one --
the self-screening by the same $3d$-electrons, which participate in the formation of other
bands due to the hybridization effects.
The fitting of screened interactions in terms
of two Kanamori parameters yields the following characteristic values of the intra-orbital Coulomb interaction
${\mathcal U}$$=$ $3.15$ ($3.19$) eV and the intraatomic exchange coupling ${\mathcal J}$$=$ $0.63$ ($0.63$) eV,
for the vanadium type V1 (V2).\cite{Kanamori}

  All parameters of the model Hamiltonian can be found in Ref.~\onlinecite{SM}. After the construction,
the model is solved in the HF approximation.\cite{downfolding}
Since the HF approximation is known to have limitations for the metallic systems, we will also discuss
possible impact of correlation interactions on the semi-quantitative level.

\begin{figure}[h!]
\begin{center}
\includegraphics[width=5cm]{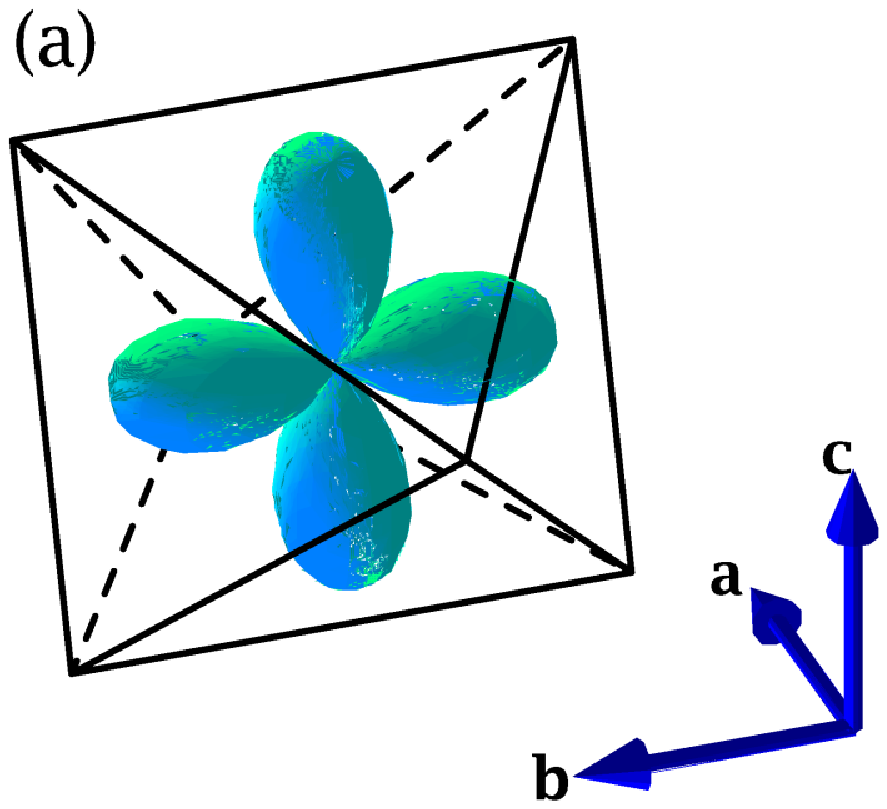}
\includegraphics[width=5cm]{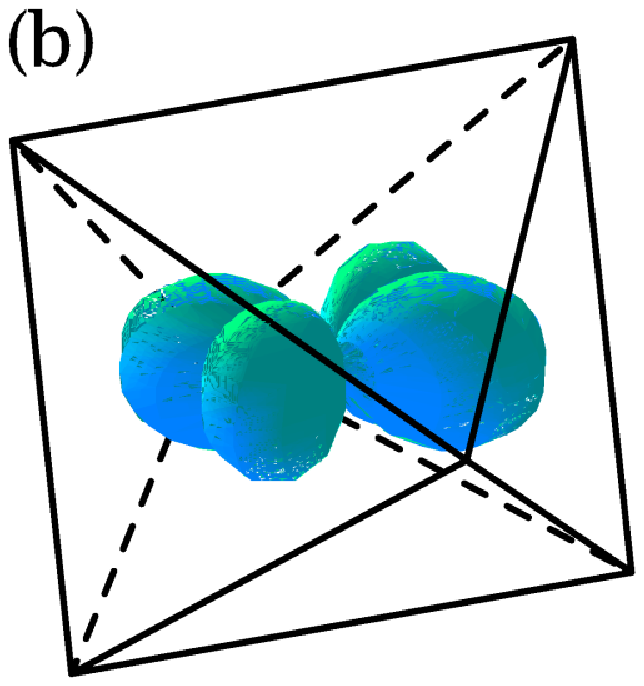}
\includegraphics[width=5cm]{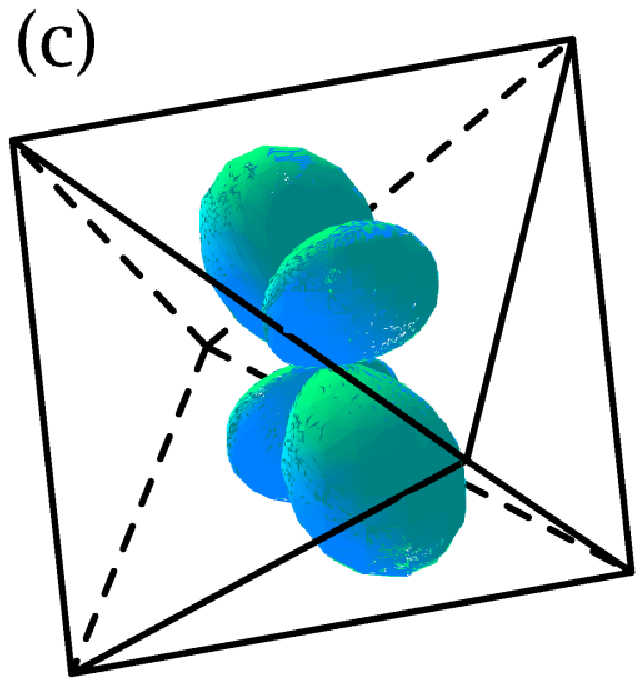}
\end{center}
\begin{center}
\includegraphics[width=5cm]{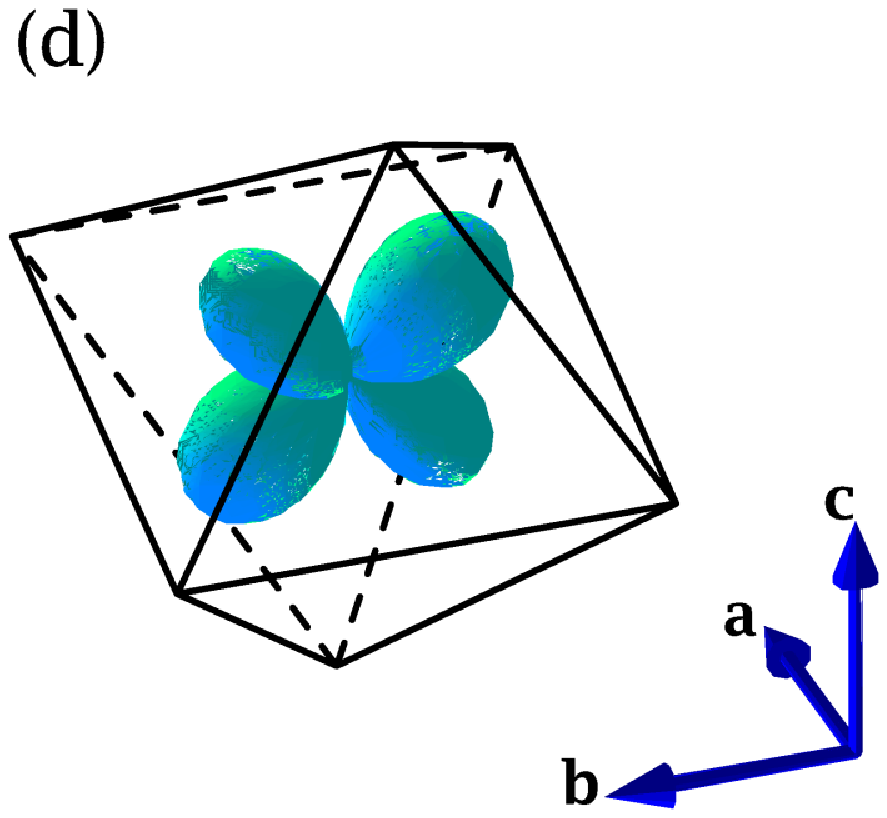}
\includegraphics[width=5cm]{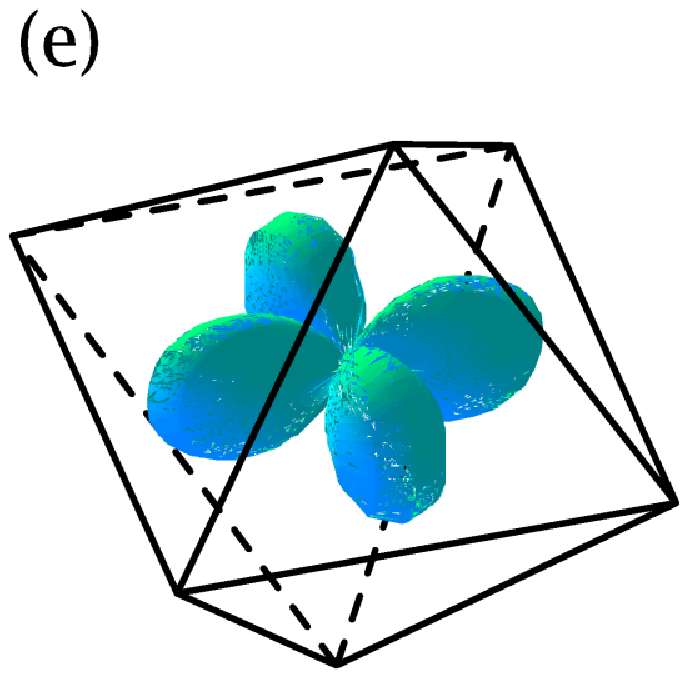}
\includegraphics[width=5cm]{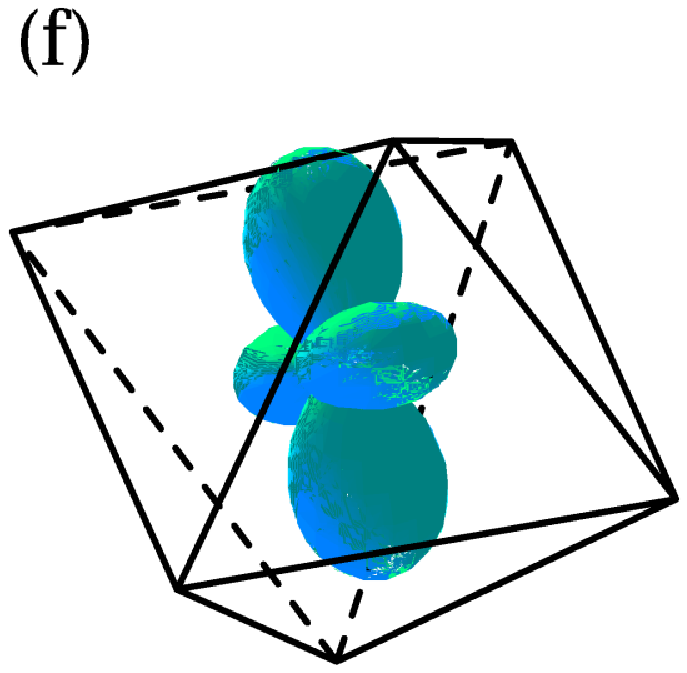}
\end{center}
\caption{\label{orbitals} (color online).
Electron density distribution, corresponding to the crystal-field orbitals
$\phi^1_{S1}$ (a), $\phi^1_A$ (b), and $\phi^1_{S2}$ (c) at the V-site 1; and
$\phi^5_A$ (d), $\phi^5_{S1}$ (e), and $\phi^5_{S2}$ (f) at the V-site 5.}
\end{figure}

  Before going into details, we would like to emphasize that the analysis of electronic and
magnetic properties of NaV$_2$O$_4$ can be greatly
simplified by considering the symmetry aspects. It appears that besides the CF splitting,
which lifts the degeneracy of atomic $t_{2g}$-levels and thus specifies subspace of the occupied orbitals,
another important factor is the \textit{symmetry} of orbitals,
which are obtained from the diagonalization of the
CF Hamiltonian at each V-site $i$. They can be schematically
denoted as $(\phi^i_{S1},\phi_A^i,\phi^i_{S2})$, for the vanadium type V1 ($i$$= 1$-$4$),
and $(\phi^i_A,\phi^i_{S1},\phi^i_{S2})$, for the
vanadium type V2 ($i$$= 5$-$8$). In the other words, these are the eigenvectors of the
CF Hamiltonian, which corresponds to the eigenvalues (the CF splitting), listed above.
In these notations, $\phi^i_S$ and $\phi^i_A$ are symmetric and antisymmetric eigenvectors
with respect to the mirror reflection $y \rightarrow -$$y$, which transform each V-site to itself
(with the additional shift by ${\bf b}/2$). For example, for the inequivalent V-sites 1 and 5 (see Fig.~\ref{afm}),
these orbitals have the following form:
$\phi^1_{S1}$$= 0.84 |3z^2$$-$$r^2 \rangle + 0.19 | zx \rangle + 0.51 |x^2$$-$$y^2 \rangle$,
$\phi^1_A$$= 0.94 |xy \rangle  - 0.33 | yz \rangle$,
$\phi^1_{S2}$$= 0.42 |3z^2$$-$$r^2 \rangle - 0.82 | zx \rangle - 0.38 |x^2$$-$$y^2 \rangle$,
$\phi^5_A$$= 0.27 |xy \rangle  - 0.96 | yz \rangle$,
$\phi^5_{S1}$$= 0.01 |3z^2$$-$$r^2 \rangle + 0.16 | zx \rangle + 0.99 |x^2$$-$$y^2 \rangle$, and
$\phi^5_{S2}$$= 0.41 |3z^2$$-$$r^2 \rangle + 0.90 | zx \rangle - 0.15 |x^2$$-$$y^2 \rangle$.
The shape of these orbitals is explained in Fig.~\ref{orbitals}.
Similar orbitals at other V-sites can be generated by applying the symmetry operations
of the space group $Pnma$. The fact that two lowest levels, which are split off by the crystal field,
belong to different representations (correspondingly, symmetric and antisymmetric one) has very
important consequences and greatly reduces the number of possible solutions of the electronic
model, which we have to consider. Indeed, due to the large orbital polarization,
which is driven by the on-site Coulomb interaction ${\mathcal U}$, the local populations of occupied
orbitals ($n$), corresponding to the $d^{1.5}$ configuration of V-sites, can be either $1$ or $1/2$.
Since these orbitals belong to different representations and, therefore, do not mix with each other,
we have to consider only four possible HF solutions, corresponding to the following populations of
the symmetric
and antisymmetric orbitals of the vanadium types V1 and V2:
$(n^{\rm V1}_S,n^{\rm V1}_A,n^{\rm V2}_S,n^{\rm V2}_A)$$=$ $(1/2,1,1/2,1)$, $(1,1/2,1/2,1)$,
$(1/2,1,1,1/2)$, and $(1,1/2,1,1/2)$. Moreover, the antisymmetric orbitals
at each site will simply coincide with $\phi_A^i$ (the only possible antisymmetric orbitals
in the $t_{2g}$ basis),
while each occupied symmetric orbital is a linear combination of $\phi^i_{S1}$ and $\phi^i_{S2}$, which is
obtained in the process of iterative solution of the HF equations. Finally,
for each orbital configuration, we should find the magnetic solution, corresponding to the
total energy minimum.

  The behavior of NN transfer integrals in the chains can be also understood, by considering
the symmetry arguments. Since for the edge-sharing geometry of the VO$_6$ octahedra, the main
contribution to the NN integrals is caused by the $dd\sigma$ interactions, which are possible only
between orbitals of the $|3z^2$$-$$r^2 \rangle$ and $|x^2$$-$$y^2 \rangle$ type, it is reasonable to expect
that the transfer integrals in the bonds $1$-$1'$ ($3$-$3'$) and $5$-$5'$ ($8$-$8'$) will be large
between symmetric orbitals, which have these components,
and small -- between antisymmetric ones, which do not have them due to the symmetry constraint.
This is indeed
roughly consistent
with the behavior of the transfer integrals, depicted in Fig.~\ref{struct}. Thus, from the viewpoint of the DE
interactions in the chains, it is more favorable energetically to make the antisymmetric orbitals
fully localized ($n$$=$$1$), and to construct the metallic band from the symmetric orbitals ($n$$=$$1/2$).
This constitutes the main idea behind results of calculations, which will be
presented in the next section.

\section{\label{Results} Results and Discussions}

\begin{figure}[h!]
\begin{center}
\includegraphics[width=7cm]{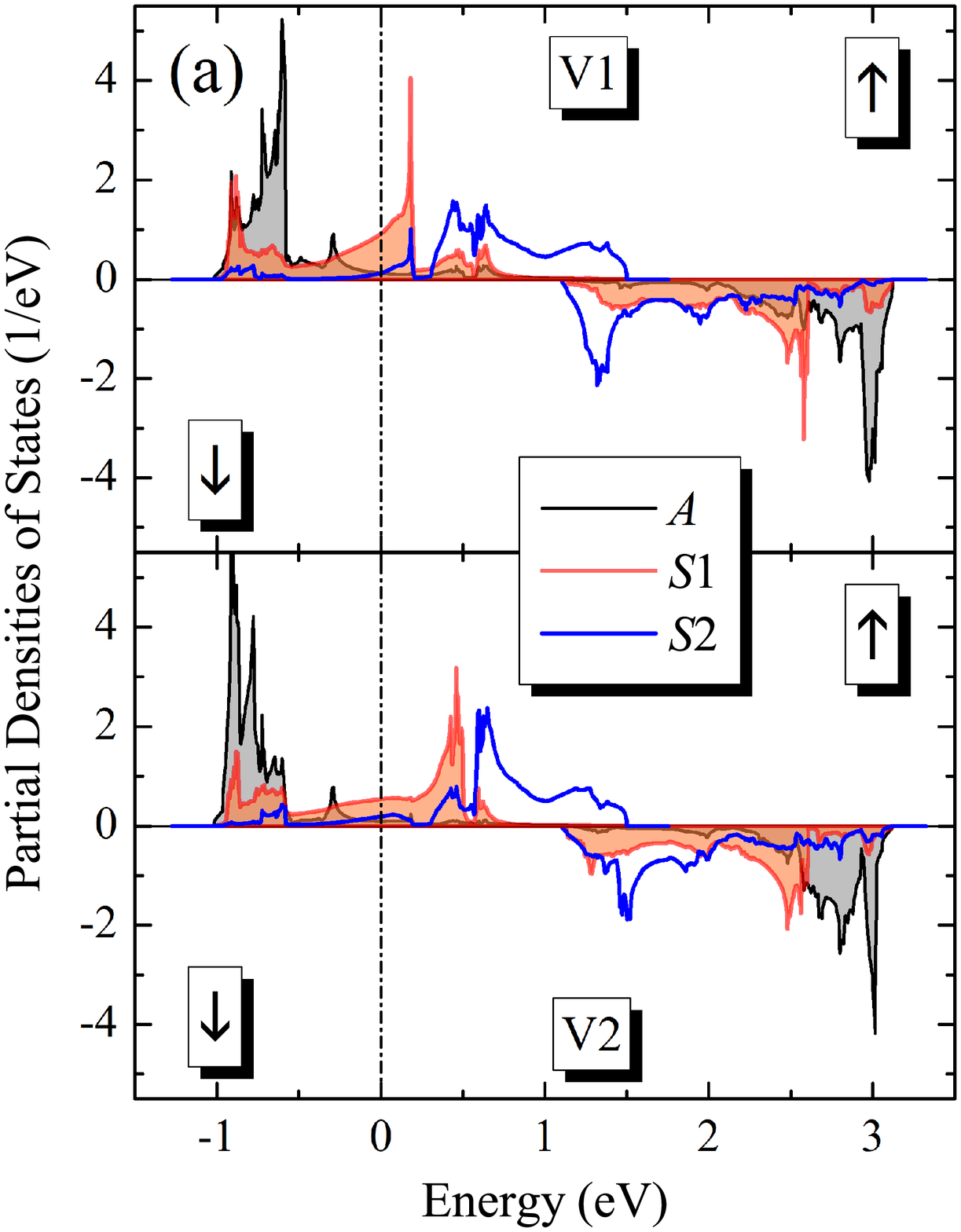}
\includegraphics[width=7cm]{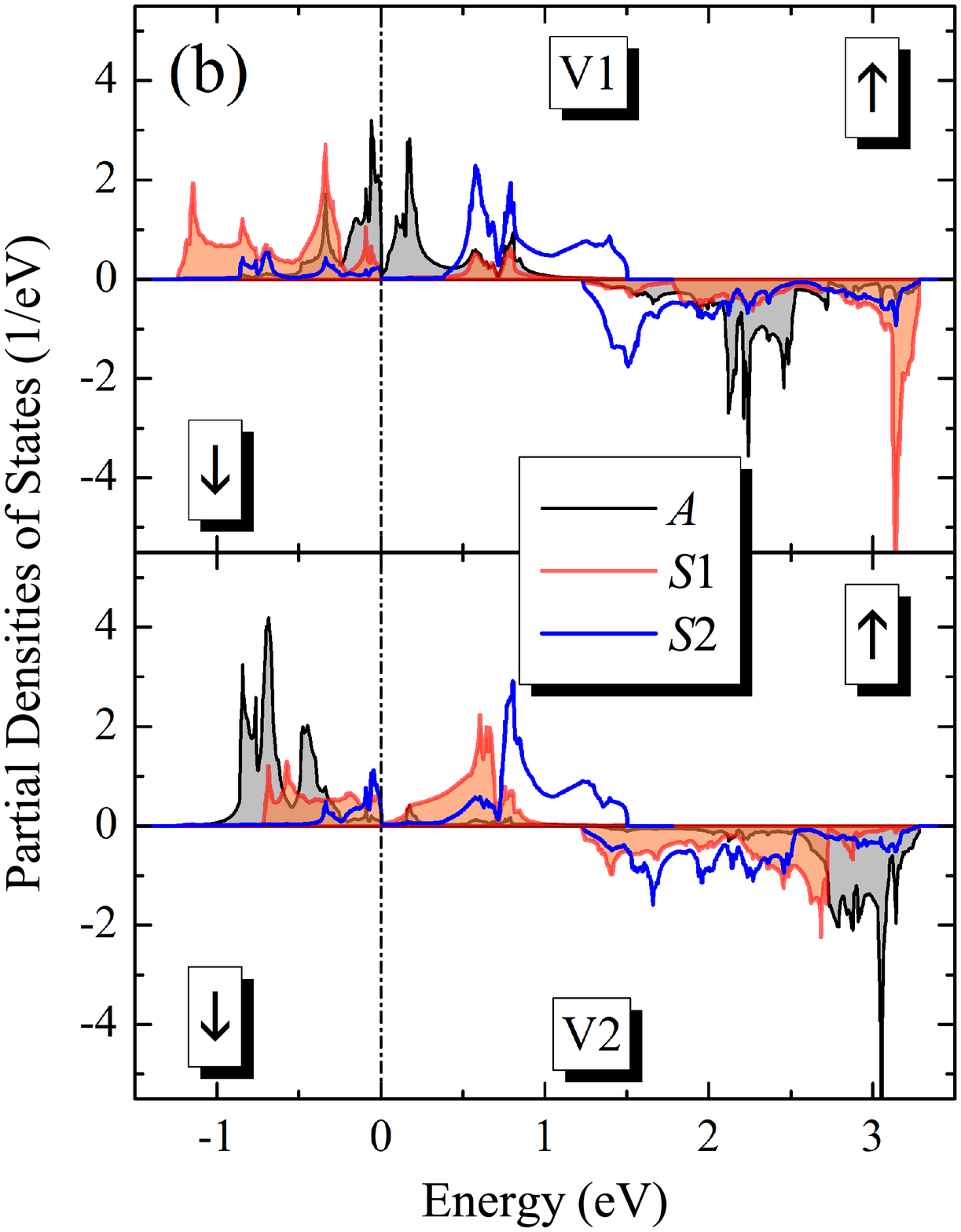}
\end{center}
\caption{\label{fig.DOS} (color online).
Partial densities of states, corresponding to the ferromagnetic spin ordering
with different orbital configurations: ${\cal O}$1 (a) and ${\cal O}$2 (b).
V1 and V2 denote two inequivalent vanadium sites, and $A$, $S1$ and $S2$ --
contributions of one antisymmetric and two symmetric orbitals, obtained from
the diagonalization of the crystal field. The Fermi level is at zero energy (shown by dashed-dotted line).}
\end{figure}

  We start with the analysis of the FM spin ordering, and attempt to stabilize different orbital structures
in the HF approximation, as described above.
By doing so, we were able to obtain only two orbital configurations,
corresponding to
$(n^{\rm V1}_S,n^{\rm V1}_A,n^{\rm V2}_S,n^{\rm V2}_A)$$=$ $(1/2,1,1/2,1)$ and $(1,1/2,1/2,1)$.
In the following, we will call them as ${\cal O}$1 and ${\cal O}$2, respectively.
Two other solutions systematically converge to one of these two.
The ${\cal O}$1 configuration appears to be lower in energy by about $9$ meV per one formula unit -- the reason
will become clear below.
The behavior of one-electron densities of states, corresponding to the orbital configurations ${\cal O}$1 and ${\cal O}$2,
is explained in Fig.~\ref{fig.DOS}.
Both solutions are half-metallic (HM).
As expected for the ${\cal O}$1 configuration, the majority ($\uparrow$)-spin $A$-states are almost fully occupied
(the small weight in the unoccupied part of the spectrum is due to the mixing
between states of the $A$- and $S$-symmetry, which is caused by transfer integrals -- see Fig.~\ref{struct}).
The metallic band is mainly formed by the $S1$-states, with the small admixture of the
$S2$-states. For the ${\cal O}$2 configuration, two V-sites display different behavior: at the site V1,
the $\uparrow$-spin bands of the $A$ and $S$-symmetry are half- and totally-filled, respectively,
while at the site V2, this filling is reversed. Moreover, the ${\cal O}$2 alignment leads to a
pseudo-gap at the Fermi level.

  The parameters of interatomic magnetic interactions are listed in Table~\ref{tab.exch}.
\begin{table}
\caption{\label{tab.exch}
Main parameters of interatomic magnetic interactions (in meV), calculated near the ferromagnetic state
for two orbital configurations ${\cal O}$1 and ${\cal O}$2. Most of the calculations have been performed for the
three-orbital model (the so-called `$t_{2g}$ model'). 
In addition, some test calculations for the
${\cal O}$1 configuration have been also performed using the five-orbital model (the so-called `$t_{2g}+e_g$ model').}
\begin{ruledtabular}
\begin{tabular}{lrrr}
  bond     & ${\cal O}$1 ($t_{2g}$$+$$e_g$) & ${\cal O}$1 ($t_{2g}$) & ${\cal O}$2 ($t_{2g}$) \\
\hline
$1$-$1'$   &    $35.9$ &    $30.1$    &      $0.5$      \\
$1$-$2$    &     $2.9$ &    $2.0$     &      $0.4$      \\
$1$-$5$    &     $9.3$ &    $7.7$     &      $7.1$      \\
$1$-$8$    &     $9.6$ &    $10.8$    &     $15.4$      \\
$1$-$1''$  &  $-$$5.8$ &    $-$$5.4$  &     $-$$0.6$    \\
$5$-$5'$   &    $37.8$ &    $30.0$    &     $26.0$      \\
$5$-$6$    &     $4.1$ &    $3.8$     &      $4.1$      \\
$5$-$5''$  &  $-$$3.2$ &    $-$$1.0$  &      $0.4$      \\
\end{tabular}
\end{ruledtabular}
\end{table}
They were calculated for two orbital configurations, by applying the perturbation theory
expansion with respect to the infinitesimal spin rotations near the FM state.\cite{Liechtenstein}
This procedure corresponds to the \textit{local} mapping of the total energy change near each
equilibrium onto the spin model
$\hat{\cal H}_S= -$$\sum_{i>j} J_{ij} {\bf e}_i {\bf e}_j$, where ${\bf e}_i$ is the \textit{direction}
of magnetic moment at the site $i$. The advantage of this procedure is that the parameters
$J_{ij}$ depend on the electronic structure of the spin-orbital state, in which they are calculated,
and can be regarded as the local probe of this state. One can clearly see that the change
of the orbital state ${\cal O}$1$\rightarrow$${\cal O}$2 has a dramatic effect on 
magnetic interactions in the V1 chains and results in the sharp drop of the the 
$J_{11'}$ interaction.
Instead, the interchain interaction $J_{18}$ tends to increase due to the
orbital reconstruction. On the other hand, the orbital configuration of V2 does not change so much.
Therefore, two solutions yield similar values of interatomic magnetic interactions $J_{55'}$ and $J_{56}$.
As a test, we have also constructed the five-orbital model ($t_{2g}$$+$$e_g$) and performed
similar calculations of interatomic magnetic interactions.
For ${\cal O}$1, these results are also shown in Table~\ref{tab.exch}. One can see that three- and five-orbital
models provide similar values of $J_{ij}$. Thus, we have confirmed that the three-orbital model captures basic
magnetic properties of NaV$_2$O$_4$ pretty well, and in the following we will focus on the analysis of 
only this model.
We have also confirmed that the change of the spin ordering for a given orbital ordering
does not significantly modify the behavior of interatomic magnetic interactions. Therefore,
in the following we will focus on the behavior of parameters obtained in the FM state.

  Let us concentrate on the low-energy configuration ${\cal O}$1 and discuss
the origin of the AFM instabilities, which were observed in the experiment.
>From the analysis of $J_{ij}$ (Table~\ref{tab.exch}), one can see that
all NN interactions are FM. Furthermore, the FM coupling within the chains ($J_{11'}$ and $J_{55'}$)
is the strongest one,
while the coupling between the chains of equivalent V-atoms
($J_{12}$ and $J_{56}$), is considerably weaker (and weaker than the interactions 1-5 
and 1-8  between different double chains).

Thus, the most probable candidate for the
AFM ground state should be the one, where the spins between the chains of equivalent
vanadium atoms are coupled antiferromagnetically and the double-chains themselves are
coupled ferromagnetically (i.e., corresponding to the
$\uparrow \downarrow \uparrow \downarrow \uparrow \downarrow \downarrow \uparrow$
spin alignment at the V-atoms 1-8, depicted in Fig.~\ref{afm}), which seems to consistent
with the experimental suggestion.\cite{yamaura_07}
In this sense, it is right to say that the quasi-one-dimensional magnetic structure, realized
in NaV$_2$O$_4$, is due to the ${\cal O}$1 orbital ordering.
Moreover, the second-neighbor interactions in the chains ($J_{11''}$ and $J_{55''}$) are AFM.
Such a behavior can be viewed as a precursor of the helical order, which was suggested
in some experimental studies.\cite{takeda_11}
Note that the AFM character of the second-neighbor interactions is
expected for the DE systems.\cite{PRL99}
Nevertheless, our calculations seem to overestimate the tendency towards the ferromagnetism.
For example, the FM character of interactions $J_{12}$ and $J_{56}$ (Table~\ref{tab.exch})
definitely favors the FM ground state, rather than the
the AFM $\uparrow \downarrow \uparrow \downarrow \uparrow \downarrow \downarrow \uparrow$ one.
The same tendency was found in the total energy calculations, where
the AFM
solution was systematically higher than the FM one
($\Delta E$$=$ $9$ and $11$ meV per one formula unit
in the three- and five-orbital model, respectively).
Moreover, since FM $J_{11'}$ ($J_{55'}$) largely exceeds the AFM $J_{11''}$ ($J_{55''}$),
the helical ordering cannot be realized either.

  We attribute these discrepancies to the HF approximation, which has
some limitations for metallic systems. Below, we qualitatively discuss the role
of correlation interactions, beyond the HF approximation, and argue that 
they can indeed resolve the problem. For these purposes,
it is convenient to decompose each NN interaction in terms of double exchange
and superexchange (SE) contributions: $J_{ij} = J_{ij}^{\rm DE}$$+$$J_{ij}^{\rm SE}$.
Formally, such a decomposition can be done for the
HM electronic structure, by assuming that the splitting between the
majority ($\uparrow$)- and minority ($\downarrow$)-spin states can be described by a single parameter $\Delta_{\rm ex}$
and expanding each $J_{ij}$ in terms of $1/ \Delta_{\rm ex}$. Thus, $\Delta_{\rm ex}$
has a meaning of (averaged) intraatomic spin splitting. Then, the zeroth-order term in this
expansion corresponds to $J_{ij}^{\rm DE}$, while the first-order term -- to $J_{ij}^{\rm SE}$.
The details can be found in Ref.~\onlinecite{NJP08}.
The parameters of such an expansion are listed in Table~\ref{tab.DESE}.
\begin{table}
\caption{\label{tab.DESE}
Results of decomposition of nearest-neighbor magnetic interactions for the orbital configuration ${\cal O}$1
in terms of double exchange ($J_{ij}^{\rm DE}$) and superexchange ($J_{ij}^{\rm SE}$). All values are in meV.
The parameter of intraatomic exchange splitting $\Delta_{\rm ex}$ was set equal to $2.8$ eV
in order to satisfy the condition $J_{ij} = J_{ij}^{\rm DE}$$+$$J_{ij}^{\rm SE}$ for the largest exchange coupling
$J_{18}$ 
between different V-types. For other bonds, this condition
is satisfied only approximately.}
\begin{ruledtabular}
\begin{tabular}{lrr}
  bond     &    $J_{ij}^{\rm DE}$       &    $J_{ij}^{\rm SE}$     \\
\hline
$1$-$1'$   &    $45.5$      &    $-18.8$   \\
$1$-$2$    &     $4.2$      &     $-2.4$   \\
$1$-$5$    &    $17.4$      &     $-9.5$   \\
$1$-$8$    &    $21.6$      &    $-10.9$   \\
$5$-$5'$   &    $44.4$      &    $-17.4$   \\
$5$-$6$    &     $9.1$      &     $-5.7$   \\
\end{tabular}
\end{ruledtabular}
\end{table}
We treat $\Delta_{\rm ex}$ as an adjustable parameter.
Then, by choosing $\Delta_{\rm ex} \approx 2.8$ eV,
the values of interatomic magnetic interactions $J_{ij}$ in Table~\ref{tab.exch}
can be approximated reasonably well by the combination $J_{ij} \approx J_{ij}^{\rm DE}$$+$$J_{ij}^{\rm SE}$
of DE and SE interactions,
listed in Table~\ref{tab.DESE}. Of course, the agreement is not perfect, mainly because
the decomposition relies on the rigid energy splitting between the $\uparrow$- and $\downarrow$-spin
states, which is an approximation. Nevertheless, $\Delta_{\rm ex} \approx 2.8$ eV
seems to be a good choice for the averaged splitting in the HF approximation (see Fig.~\ref{fig.DOS}).
As expected, all DE interactions are FM, while SE interactions are AFM.
The large FM coupling within the chains is stabilized by the ${\cal O}$1 orbital ordering,
which maximizes the DE interactions
(for comparison, the ${\cal O}$2 orbital arrangement yields $J_{11'}^{\rm DE}$$= 25.5$ meV and
$J_{11'}^{\rm SE}$$= -$$26.5$ meV).

  Then, it is reasonable to expect that $\Delta_{\rm ex}$ will be screened by correlation interactions.
For example, such an effect is well known in the theory of the homogeneous electron gas.\cite{BH}
Similar tendency was found in the total energy calculations
for the low-energy model, which was constructed for the series of
HM systems: the correlation interactions, treated in 
RPA, systematically decrease the
effective intraatomic spin splitting and, in a number of cases,
make this HM state unstable.\cite{condmat}
Thus, it is clear that the reduction of $\Delta_{\rm ex}$, caused by the metallic screening effects, will strengthen
$J_{ij}^{\rm SE}$, which is proportional to $1/\Delta_{\rm ex}$, and shift the balance of magnetic interactions
towards the AFM coupling (of course, provided that
the orbital configuration will not change). Particularly, in order to make the bonds $1$-$2$ and
$5$-$6$ antiferromagnetic (see Table~\ref{tab.DESE}), the spin splitting $\Delta_{\rm ex}$ should be
reduced by about 40\%. Then, if we adopt the same scaling for $J_{11'}^{\rm SE}$ and note that
$J_{11'}^{\rm DE}$ does not depend on $\Delta_{\rm ex}$, the `screened' NN interaction
$\bar{J}_{11'}$ can be estimated as $\bar{J}_{11'} \approx J_{11'}^{\rm DE}$$+$$\frac{5}{3}J_{11'}^{\rm SE} = 14.2$ meV.
In this case, one can also expect the formation of the helical magnetic state in the chain V1,
which takes place if $\bar{J}_{11'} < -$$4 J_{11''}$. Thus, the
experimental magnetic ordering is closely related to the ${\cal O}$1 orbital ordering:
at least on the semi-quantitative level, main details of the experimental magnetic structure are reflected in the
behavior of interatomic magnetic interactions.
We expect that the quantitative agreement with the experimental data could be obtained
if one goes beyond the HF approximation and consider rigorously the correlation interactions.
This is qualitatively consistent with calculations of the correlation energy in RPA,
by starting from one-electron eigenvalues and eigenfunctions, obtained in the HF approximation.
This procedure yields the following values of the total (i.e., HF plus correlation)
energy differences between AFM and FM states:
$\Delta E$$=$ $6$ and $2$ per one formula unit
for the three- and five-orbital model, respectively.
Thus, already the `single-shot calculations'
of correlation energy substantially reduce $\Delta E$ (especially for the five-orbital model).
We hope that a better agreement can be obtained by treating the correlation effects
self-consistently.\cite{condmat}

  Finally, let us discuss the anisotropy of electrical resistivity.
The $dc$ conductivity can be evaluated using the Boltzmann's equation approach
(in Rydberg atomic units):\cite{AshcroftMermin}
$$
\sigma_{\alpha \beta} = \frac{1}{2 \pi^3} \sum_l \int_{\rm BZ} d {\bf k} \tau_l ({\bf k})
v_l^\alpha ({\bf k}) v_l^\beta ({\bf k}) \left( - \frac{\partial f(\epsilon)}{\partial \epsilon}
\right)_{\epsilon = \epsilon_l ({\bf k})},
$$
where
$\epsilon_l ({\bf k})$ is the band dispersion ($l$ being the band index);
$v_l^{\alpha (\beta)} ({\bf k}) = \partial \epsilon_l ({\bf k}) / \partial k_{\alpha  (\beta)}$ is the group velocity;
$\alpha (\beta)$$=$ $x$, $y$, or $z$ in the orthorhombic frame;
$\tau_l ({\bf k})$ is the quasiparticle lifetime; and
$f(\epsilon)$ is the Fermi distribution function. For the orthorhombic $Pnma$ symmetry, this tensor is diagonal
($\sigma_{\alpha \beta} = \sigma_{\alpha \alpha} \delta_{\alpha \beta}$), and the
components of the resistivity tensor are given simply by $\rho_{\alpha \alpha} = 1/ \sigma_{\alpha \alpha}$.
Then, for a constant $\tau$, one can easily evaluate
the rations $\rho_{\alpha \alpha}/\rho_{\beta \beta}$ between different components 
and compare it with the experimental data.

  Let us start our analysis with the HM FM state.
Then, for the ${\cal O}$1 orbital ordering, we obtain the following ratios:
$\rho_{xx}/\rho_{yy} = 34$ and $\rho_{zz}/\rho_{yy} = 19$. Thus, the resistivity is highly
anisotropic, where two components perpendicular to the chains
$\rho_{xx} \approx \rho_{zz} \equiv \rho_\perp$ are about 19-34 times lager than
$\rho_{yy} \equiv \rho_\parallel$ in the chain. Therefore, the experimentally observed
anisotropy of resistivity ($\rho_\perp/\rho_\parallel > 20$, Ref.~\onlinecite{yamaura_07})
can be understood already from the viewpoint
of the ${\cal O}$1 orbital ordering, even without invoking the anisotropic
$\uparrow \downarrow \uparrow \downarrow \uparrow \downarrow \downarrow \uparrow$ AFM arrangement.

  The orbital ordering plays a crucial role in the observed anisotropy of the
resistivity tensor. For example, for the ${\cal O}$2 orbital ordering, the tensor
$\hat{\rho} = \| \rho_{\alpha \beta} \|$ is nearly isotropic
($\rho_{xx}/\rho_{yy} \approx \rho_{zz}/\rho_{yy} \approx 1$).
Similar behavior is found in the local-spin-density approximation
for the HM FM state ($\rho_{xx}/\rho_{yy} = 3$ and $\rho_{zz}/\rho_{yy} = 4$).

  As expected,\cite{yamaura_07}
the AFM $\uparrow \downarrow \uparrow \downarrow \uparrow \downarrow \downarrow \uparrow$
spin alignment additionally enhances the anisotropy of the resistivity tensor, yielding
for the ${\cal O}$1 orbital ordering
$\rho_{xx}/\rho_{yy} = 46$ and $\rho_{zz}/\rho_{yy} = 210$.
Therefore, one may ask which ordering is more important for the anisotropy of
resistivity: spin or orbital one?
The increase of $\rho_{xx}/\rho_{yy}$, in comparison with the FM alignment
(for the same orbital ordering ${\cal O}$1), is quite modest. However,
the ratio $\rho_{zz}/\rho_{yy}$ changes by factor 10.
Therefore, it is tempted to conclude that the
antiferromagnetism plays at least the same role in the anisotropy 
of the electrical resistivity as the orbital ordering.
Nevertheless, it should be also remembered that the AFM order itself is caused by the orbital
order (or, at least these two orders occur concomitantly). Thus, it is probably right to say
that this is the orbital ordering, which has twofold effect on the anisotropy of 
the electrical resistivity:
direct, which is observed already in the FM state, and indirect, via formation of the
anisotropic AFM structure.

\section{\label{summary} Summary}

  On the basis of the first-principles electronic structure calculations, we have derived
an effective low-energy model for NaV$_2$O$_4$ and employed this model for the 
analysis of electronic and magnetic properties of this material. 
The obtained transfer integrals are basically three-dimensional and by themselves do not reproduce
the quasi-1D character
of the electrical resistivity of NaV$_2$O$_4$. 
An additional important ingredient, which yields the anisotropy of electronic 
and magnetic properties should be the orbital ordering, realized in the
metallic state of NaV$_2$O$_4$. We have argued that the symmetry arguments greatly
simplify the analysis of the possible orbital states in NaV$_2$O$_4$,
and the proposed orbital ordering is one of the few candidates, which respects
the orthorhombic $Pnma$ symmetry. 
The AFM spin arrangement additionally increases the anisotropy of electrical resistivity.
However, the anisotropy caused by the orbital ordering is comparable with the
spin contribution and is formally sufficient for reproducing the experimental
value $\rho_\perp$/$\rho_\parallel>$20.\cite{yamaura_07}
Moreover, the AFM spin ordering itself is driven by the orbital ordering, which
increases the FM character of interactions in the chains and weakens the
interactions between the chains. The sizable second-neighbor AFM
interactions in the chains can be also responsible for the formation of the
helical magnetic structure, which was proposed recently.\cite{muon}
Nevertheless, within the HF approximation, the FM solution
was found to be slightly lower in energy than the AFM one. 
In order to resolve this discrepancy, 
we provided semi-quantitative arguments 
and argued that the correlation interactions, beyond the HF approximation, 
should probably reverse this tendency.

\begin{acknowledgments}
We would like to thank D. Hirai, T. Takayama, and H. Takagi
for providing us the structure parameters of NaV$_2$O$_4$ prior to the publication.\cite{takagi}
RA acknowledge the financial support from JST-PRESTO.
The work of ZVP is partly supported by RFFI-10-02-96011.
\end{acknowledgments}


\begin{thebibliography}{10}
\bibitem{remark2} Very similar parameters have been reported in Ref.~\onlinecite{yamaura_07}.
We have also confirmed that the use of the $T$$=$$130$ K experimental structure parameters
yields essentially the same theoretical results.
\end{thebibliography}

\end{document}